\documentclass[final,5p,times,twocolumn]{elsarticle}

\usepackage{amssymb}
\usepackage{amsmath}
\usepackage{graphicx} 
\usepackage{epsfig}
\usepackage{color}
\usepackage{bm}
\usepackage{cases}
\usepackage{makecell} 

\usepackage{lipsum}

\newcommand{\kbt}{k_BT}
\newcommand{\fij}{F_{ij}}

\newcommand{\rbij}{{\bf r'}_{ij}}

\newcommand{\Aij}{A_{ij}}
\newcommand{\Bij}{B_{ij}}
\newcommand{\rhoi}{\rho_{i}}
\newcommand{\rhoj}{\rho_{j}}

\newcommand{\vb}{{\bf v}}

\newcommand{\eij}{{{\bf e}_{ij}}}

\newcommand{\vcon}{{\bf{v}}}
\newcommand{\vmes}{{\bf v}'}

\newcommand{\fijcon}{{F}_{IJ}}

\newcommand{\rbijcon}{{\bf r}_{IJ}}
\newcommand{\ebijcon}{{\bf e}_{IJ}}

\newcommand{\facon}{a}
\newcommand{\fbcon}{b}

\newcommand{\wbbarij}{{\bf \overline{W}}_{ij}}
\newcommand{\wbij}{{\bf {W}}_{ij}}
\newcommand{\fa}{a'}
\newcommand{\fb}{b'}

\newcommand{\der}{\text{d}}

\newcommand{\ac}{\textbf{G}}
\newcommand{\sol}{\textbf{S}}

\definecolor{newgreen}{RGB}{1,129,30}

\usepackage{amsthm}
\biboptions{sort&compress}

\journal{...}

\begin{document}

\begin{frontmatter}



\title {Towards the Multiscale Design of Pressure Sensitive Adhesives}


\author[inst1]{Nicolas Moreno\corref{cor1}}
\cortext[cor1]{nmoreno@bcamath.org}
\affiliation[inst1]{organization={Basque Center for Applied Mathematics},
            addressline={Alameda de Mazarredo 14}, 
            city={Bilbao},
            postcode={48009}, 
            state={Basque Country},
            country={Spain}}

\author[inst1]{Elnaz Zohravi}
\author[inst2]{Shaghayegh Hamzehlou}

\affiliation[inst2]{organization={Polymat and Kimika Aplikatua Saila, Kimika Zientzien Fakultatea, University of the Basque Country UPV/EHU},
            addressline={Joxe Mari
Korta Zentroa, Tolosa Hiribidea 72}, 
            city={Donostia-San Sebastian},
            postcode={E-20018}, 
            country={Spain}}

\author[inst1]{Edgar Pati\~{n}o-Nari\~{n}o}
\author[inst2]{Malavika Raj}
\author[inst2]{Mercedes Fernandez}
\author[inst2,inst3]{Nicholas Ballard}
\author[inst2]{Jose M. Asua}
\author[inst1,inst3,inst4]{Marco Ellero,\corref{cor2}} 

\affiliation[inst3]{organization={Ikerbasque, Basque Foundation for Science},
            addressline={ Calle de Maria Diaz de Haro 3}, 
            city={Bilbao},
            postcode={ 48013}, 
            country={Spain}}
\affiliation[inst4]{organization={Complex Fluids Research Group, Department of Chemical Engineering, Faculty of Science and Engineering, Swansea University},
            addressline={ SA1 8EN}, 
            city={Swansea},
            country={United Kingdom}}

\begin{abstract}

Pressure-sensitive adhesives (PSAs) are soft polymeric materials that exhibit complex rheological and mechanical behavior governed by the interplay between polymer architecture, crosslink density, and entanglement constraints. Predicting their rheological properties from underlying microstructure remains a central challenge in adhesive design. In this work, we adopt a multiscale computational framework based on the Lagrangian Heterogeneous Multiscale Method (LHMM), coupling a macroscopic continuum description with a mesoscale polymer network model featuring breakable bonds embedded in a viscous medium. The approach enables consistent information transfer across scales and captures both elastic network response and viscous dissipation. The framework is calibrated using experimental rheological data and tensile measurements for four PSA formulations with varying gel fractions and crosslink densities. The simulations reproduce key experimental trends in storage modulus ($G'$), loss modulus ($G^{\prime\prime}$), and tensile stress–strain behavior under planar extension, while differentiating the distinct mechanical signatures of each formulation. The results elucidate how crosslink density and effective network connectivity control stiffness, stress localization, and failure characteristics. Overall, the proposed multiscale methodology provides a predictive platform for linking microstructural design parameters to macroscopic mechanical properties and offers a rational basis for the formulation and optimization of next-generation PSAs.
\end{abstract}

\begin{graphicalabstract}
\includegraphics{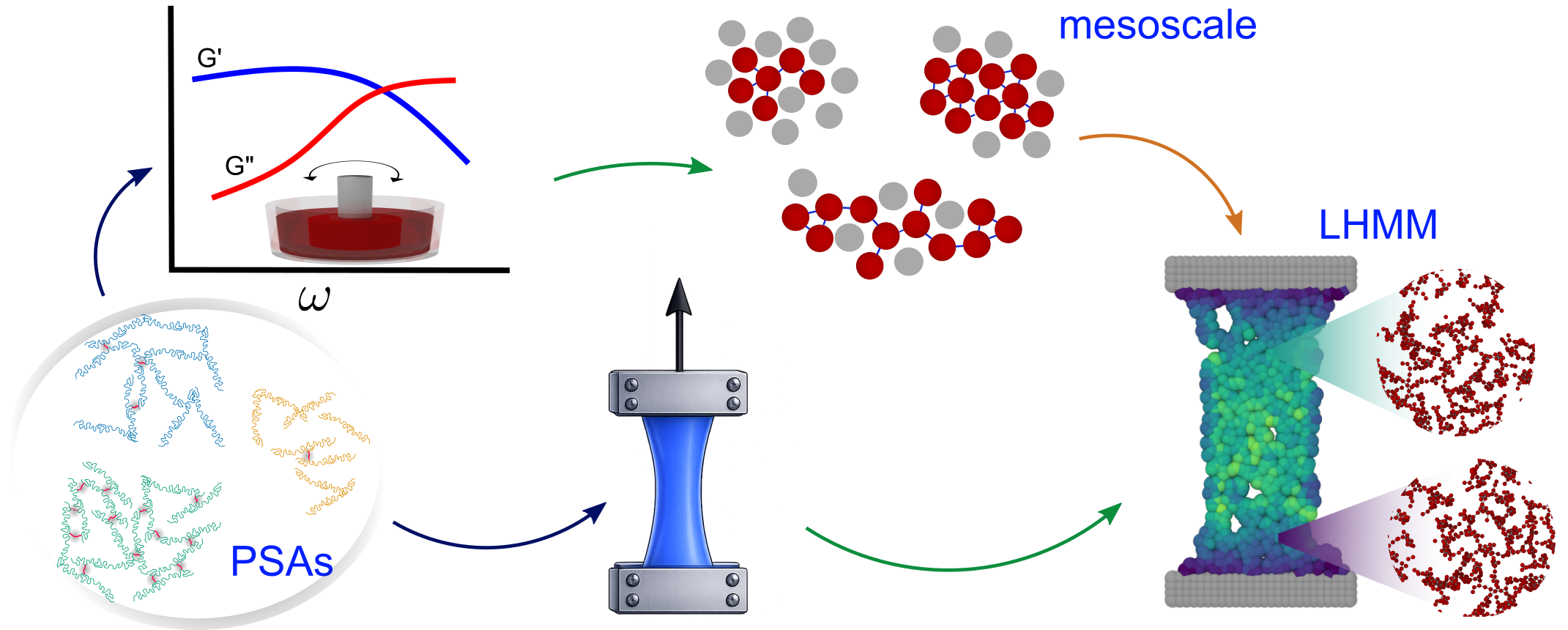}
\end{graphicalabstract}



\end{frontmatter}


\section{Introduction}
\label{sec:Introduction}

Pressure-sensitive adhesives (PSAs) are soft, viscoelastic materials that can adhere to surfaces with just a small amount of applied pressure\cite{Creton2003}. While their use is widespread and their function may seem straightforward, the adhesive-substrate interactions are in fact highly complex and are governed by a range of interrelated factors \cite{Creton2016, Raos2021}. At the molecular level, bonding typically occurs through physical interactions, whereas the overall adhesive performance is largely determined by the chain dynamics of the macromolecular structure \cite{Wang2020}. Understanding these mechanisms is essential for the development of next-generation adhesive materials with tailored properties.
\\\\
From an experimental point of view, it has been widely demonstrated that optimizing adhesive behavior requires a delicate balance in rheological properties, which can be largely controlled through the macromolecular structure of the polymer that makes up the PSA \cite{Ballard2024, Jovanovi2004}. For example, when the polymer chains in the material are heavily crosslinked, they tend to display almost purely elastic behavior with a high shear modulus, which leads to relatively good resistance to shear \cite{Tobing2001, Arrowood2023, Zosel1994}. However, the tack, which is commonly measured by measuring the energy required to remove a probe that has been brought into contact with the adhesive under controlled conditions \cite{Crosby1999, Lakrout1999}, and the peel strength of heavily crosslinked PSAs is typically poor. In contrast, for lower molecular weight, linear polymers, the adhesive is more liquid-like and can show relatively good tack and peel but has effectively no resistance to shear. To make matters more complicated, the adhesive properties strongly depend on the non-linear rheological properties of the polymer \cite{Creton2001, Creton2009, Chopin2018}, which are again heavily influenced by the macromolecular structure of the polymer. Optimizing a PSA thus involves a complicated balancing act between the elastic and viscous nature of the polymer, which can be achieved through control of polymer composition (including addition of function monomers that enhance surface interactions) \cite{Gower2006, Gower2004}, and the macromolecular structure of the polymer \cite{Lakrout2001, Tobing2001b, Tobing2001c, Deplace2009, Bratasanu2025}. 
\\\\
To aid in the design of PSAs, there have been numerous attempts to model the adhesive behavior of polymer materials \cite{Creton2000, Gay1999, Chikina2000, Yamaguchi2006, Foteinopoulou2006, Glassmaker2008, Yamaguchi2018, Varchanis2021, Sgouros2025}. Modeling approaches can use molecular models based on kinetic theory, which provide insights into the relationship between molecular structure and rheological properties \cite{Bird2016}. At the mesoscale, Sgouros et al. \cite{Sgouros2025} modeled polymeric adhesives adopting a systematic parameterization from atomistic simulations or experimental data to capture polymer chain dynamics and entanglement effects through entropic springs and slip-springs. The resulting dynamics were governed by Langevin equations coupled with kinetic Monte Carlo treatments of slip-spring evolution, allowing physically informed prediction of adhesive deformation behavior at spatial and temporal scales close to the polymer coil dimension. Other approaches typically describe macroscopically the stress tensor through continuum polymeric-constitutive equations \cite{Larson2015,Varchanis2021}. For example, finite element method simulations have been used to study the stretching Newtonian \cite{Foteinopoulou2006} and viscoelastic \cite{Varchanis2021} filaments with preexistent cavities. These numerical approaches used polymeric-constitutive equations to reproduce the rheological response of the adhesives, and have provided an important understanding on the growth of cavities both in the bulk and at the adhesives-substrate interface. However, the parameters of the constitutive models may not be straightforward to correlate with microstructural features (such as the degree of crosslinking) requiring the additional exploration of different models. Yamaguchi et al. \cite{Yamaguchi2018} have discussed a relatively simple probe tack experiment based on a phenomenological analysis of the debonding process and cavity expansion behavior in which the probe tack curve can be predicted. The model was in qualitative agreement with experimental results and was able to provide insights into how material properties (i.e shear modulus) will influence the tack energy, but it makes a number of simplifications that make quantitative predictions challenging. Thus, despite the advances made in understanding the response of PSAs, the multiscale mechanisms that are operating are of such complexity that directly modeling the adhesive response based on macromolecular structure remains a challenge.

This work aims to develop additional multiscale tools that are capable of taking into account the complexity of macro and microscale parameters that determine the rheological behavior of PSAs. Multiscale modeling has emerged as a promising approach to gain insights into fundamental mechanisms and guide the design of novel materials. With advancements in numerical methods and high-performance computing systems, developing robust and tractable multiscale models for investigating complex polymeric materials is now within reach. This includes multiscale methodologies (using Eulerian \cite{Moreno2013,E2007} and Lagrangian \cite{moreno2023generalized} discretizations) and model reduction techniques that allow a one-to-one mapping with physical systems on relevant temporal scales for practical applications\cite{Moreno2014, Moreno2015}.

\begin{figure*}[!htbp]
    \centering
    \includegraphics[width=0.9\linewidth]{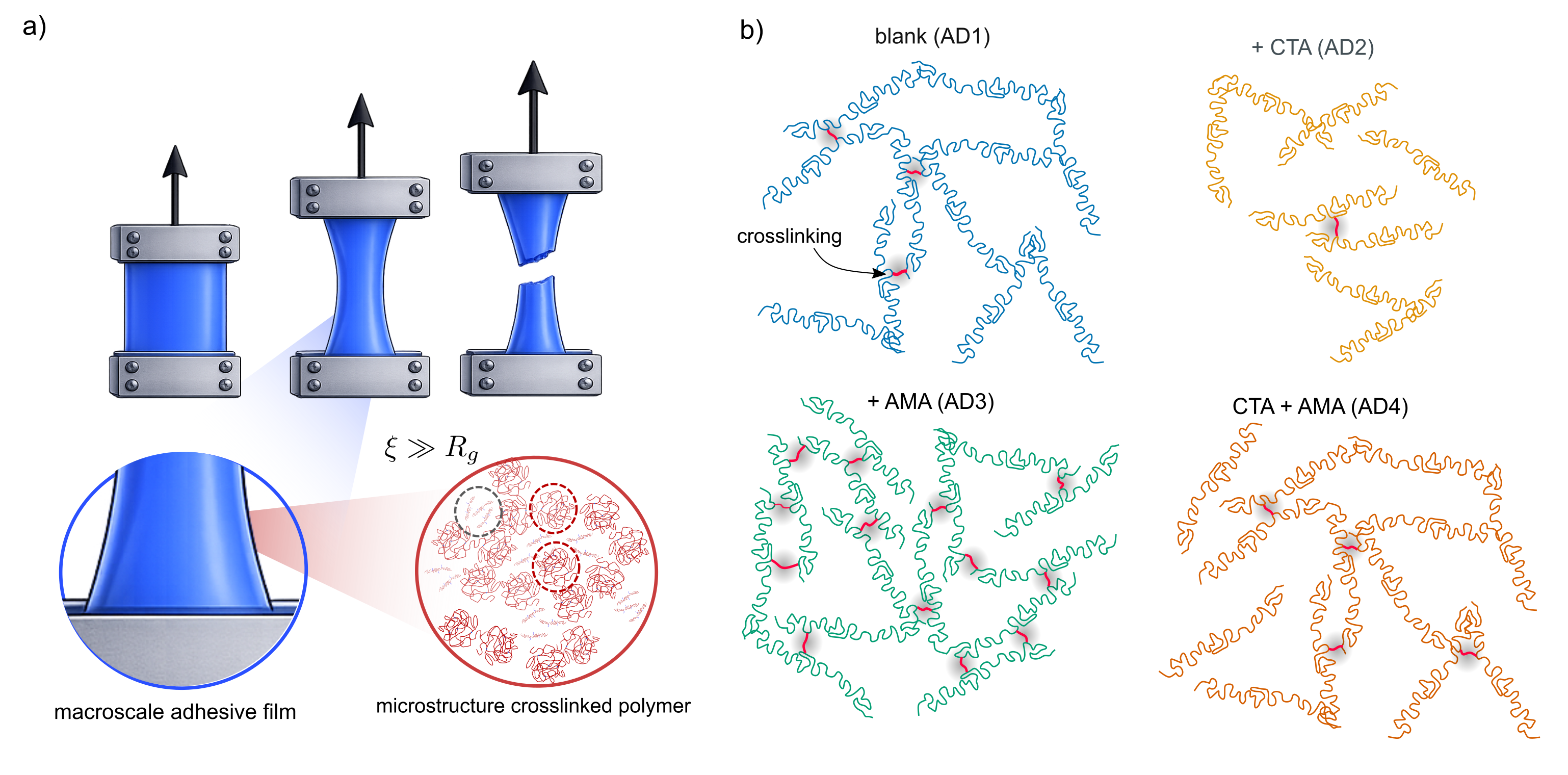}
    \caption{Schematic of multiscale features of tensile test experiments and adhesives manufactured. a) Schematic of the tensile test experiments. The sample is deformed at a constant elongation rate, while the stress is recorded as a function of strain. At the microscale, the properties of the polymeric network (e.g. degree of crosslinking, molecular weight distribution, etc) will determine the response of the material to the deformation, and thus the macroscopic stress-strain response. b) The four samples (AD1 to AD4) were synthesized with varying amounts of chain transfer agent (CTA) and crosslinker (AMA). Addition of CTA leads a reduction in the overall molecular weight of the polymer compared to AD1, and overall degree of crosslinking, whereas addition of AMA favors the crosslinking of the polymer chains, while keeping the length of the polymer chains relatively unchanged compared to the blank sample (AD1). Mixtures of CTA and AMA (AD4) lead to microstructural features that are intermediate between the two extremes, depending on the formulation.}
\label{fig:expSketch}
\end{figure*}

In this work, we adopt a heterogeneous multiscale modelling approach \cite{moreno2023generalized} to investigate the response of PSAs with varying degrees of crosslinking. We consider a two-way coupling between macro and microscales, where the polymeric network features are modelled at the microscale, and the bulk behavior of the material is described at the macroscale. Microscale models are initially calibrated to reproduce the experimental rheological and structural features of the different adhesive formulations. Then, we use the calibrated microscale models to conduct multiscale simulations of tensile tests to characterize the macroscopic response of the material, and compare the results with experimental measurements. Tensile test are intrisically multiscale, as illustrated in Figure \ref{fig:expSketch}.a, where the macroscopic response of the material is determined by the underlying microstructural features of the polymeric microstructure, and the deformation process can lead to the emergence of different physical phenomena at different scales. Thus, tensile tests provide an ideal platform to investigate the multiscale response of PSAs and to validate the proposed multiscale modeling approach. Our simulations show that the multiscale models can provide valuable insights into the mechanisms governing the material response across scales. Due to the multiplicity of physical phenomena (polymer coil dynamics, polymer-substrate interaction, cavitation, etc) involved in the adhesion process, herein, the stress-strain response of the material was evaluated using a uniaxial tension test. Cavity formation, fibrillation, and the effects of polymer-substrate interactions will be addressed in future publications. 

\section*{Methods \& Models}

\subsection*{Experimental synthesis and characterization of adhesives}

We synthesized the adhesives using a seeded semi-continuous emulsion polymerization process. The first stage involved preparing a poly(n-butyl-acrylate) (PBA) seed in a batch setup (see S.Table S.1, the initial seed formulation), followed by a second stage conducted in a glass reactor with controlled feeding of monomer pre-emulsion and initiator solutions over three hours at 75°C.  A detailed description of the materials, setup, and procedure is provided in Supplementary Data S1. 
\\\\
Four adhesive samples (AD1 to AD4) were prepared by varying the amounts of chain transfer agent (n-dodecyl mercaptan, CTA) and crosslinker (allyl methacrylate, AMA).In Table \ref{tab:ad_formulation}, the differences in the content of CTA and AMA among samples are summarized. The formulation used for the preparation of the "blank" sample without any CTA or AMA (referred to as AD1) is provided in S.Table S.2.  In Figure \ref{fig:expSketch}.b, we illustrate the differences among samples. Compared to the blank sample (AD1), the addition of CTA can lead to the formation of shorter polymer chains reducing the molecular weight, and overall degree of crosslinking. In contrast, the addition of AMA favors the crosslinking of the polymer chains (network connectivity), while keeping the length of the polymer chains relatively unchanged compared to AD1. Mixtures of CTA and AMA (AD4) allows us to customize the adhesive response by tuning the microstructural features of the polymeric network, which are intermediate between the two extremes, depending on the formulation. The different adhesives samples were characterized using several methodologies: \textit{i)} molar mass distribution; \textit{ii)} gel content;  \textit{iii)} rheological response; and \textit{iv)} tensile tests. The complete description of the different tests, setup, and parameters are provided in the Supplementary Data S1.
\begin{table}[htbp!]
\centering
\caption{Summary of formulation differences among the four synthesized adhesive samples}
\begin{tabular}{|c|c|c|c|}
\hline
\textbf{Sample} & \textbf{CTA (g)} & \textbf{AMA (g)} & \makecell{\textbf{Gel Content} \\ \textbf{(\%)}} \\ \hline
\makecell{AD1 \\ (blank)} & 0 & 0 & 69 \\ \hline
\makecell{AD2 \\ (+CTA)} & 1.64 & 0 & 15 \\ \hline
\makecell{AD3 \\ (+AMA)} & 0 & 0.63 & 94 \\ \hline
\makecell{AD4 \\ (+CTA+AMA)} & 0.41 & 0.35 & 69 \\ \hline
\end{tabular}
\label{tab:ad_formulation}
\end{table}

The molar-mass distribution of each of the samples was analyzed using asymmetric-flow field-flow fractionation (AF4) coupled with multi-angle light scattering (MALS) and refractive index (RI) detectors (see Supplementary Data  S1.2).  The gel content of the adhesives was determined as the fraction of polymer remaining insoluble after Soxhlet extraction with THF. The procedure involved weighing the sample before and after extraction and calculating the gel content based on weight differences (see Supplementary Data S1.1). In Table \ref{tab:ad_formulation}, we provide a summary of the gel content for each sample.
\\\\
We conducted the rheological characterization of the adhesives using a strain-controlled rheometer (ARES G2, TA Instruments) to measure the storage modulus ($G'$) and loss modulus ($G''$) under oscillatory shear (via Small Amplitude Oscillatory Shear -SAOS- measurements)(see Supplementary Data S1.3). To streamline the comparison between the experimental and simulation results, we normalize $G^{\prime}$ and $G^{\prime\prime}$ with the maximum value of $G^{\prime}$ measured for sample AD1 (blank). The angular frequency $\omega$ was also normalized using the highest experimental frequency applied. This normalization allows the identification of the relative change of $G^{\prime}$ and $G^{\prime\prime}$ among samples. 
\\\\
Room-temperature tensile stress-strain measurements were performed using an ARES-G2 rheometer equipped with tension geometry. Film specimens with a width of $d_0\simeq5$mm and a thickness of $\simeq0.4$mm were mounted with an initial clamp-to-clamp distance (gauge length) of $\simeq5$mm and deformed at a constant elongation rate of $0.01$mm s$^{-1}$, while stress was recorded as a function of strain.

\subsection*{Computational Multiscale Model}

\begin{figure}[!htbp]
\centering
\includegraphics[width=0.9\linewidth]{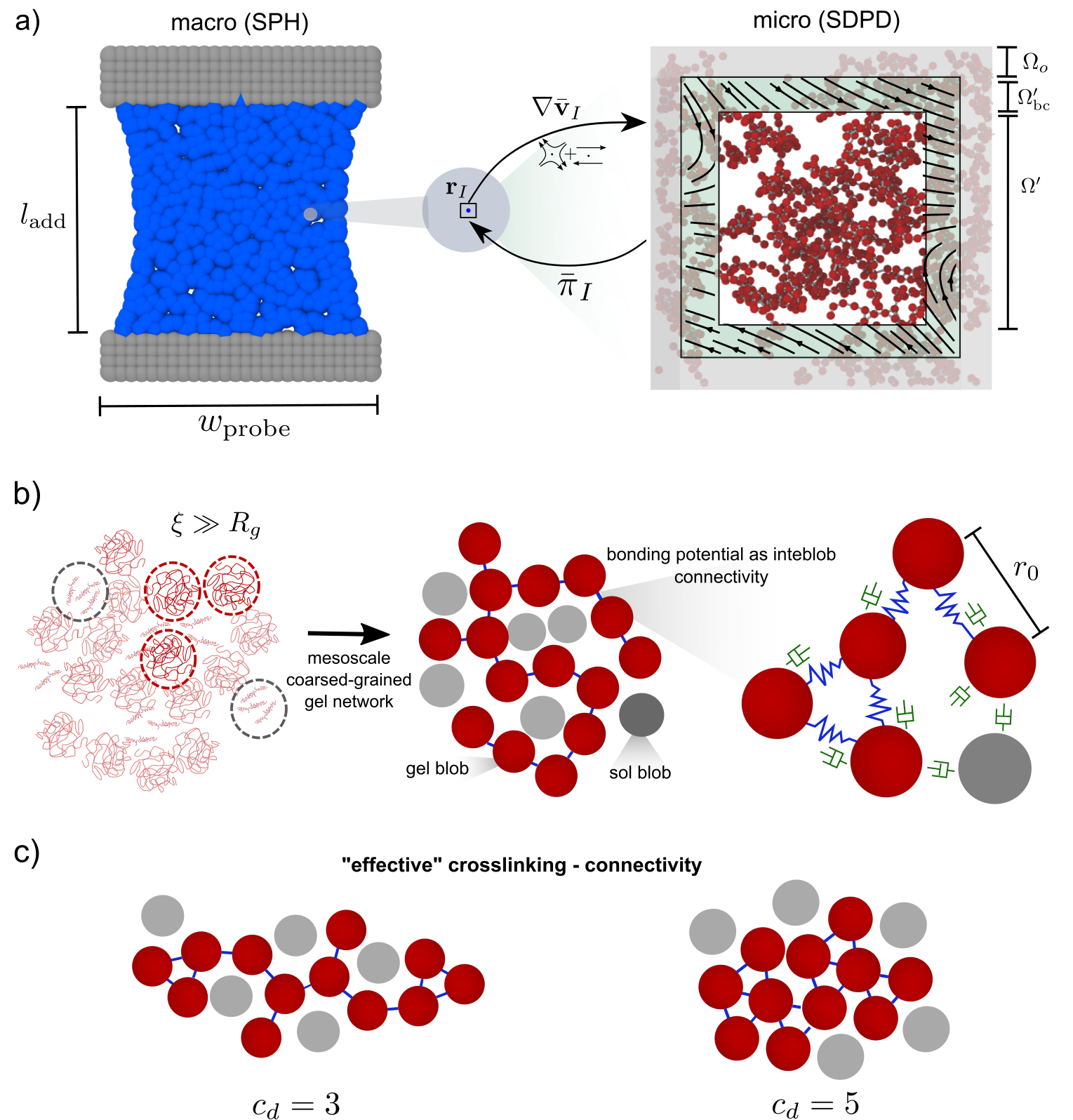}
\caption{a) LHMM sketch, b) Schematic mapping approach. The polymeric network is discretized as interconnected gel blobs (red) and low molecular weight fractions as sol blobs (grey). The gel volumetric fraction determines the total amount of gel blobs in the system. Stretchable bonding potentials (blue springs) are used to capture network connectivity, whereas all the blobs (gel and sol) in the system interact \textit{via} viscous forces (green dashpot); c) Illustration of the cross-linking degree for two different systems with $c_d=3$ and $c_d=5$.}
\label{fgr:modelSketch}
\end{figure}

We used the recently proposed Lagrangian heterogeneous multiscale method (LHMM)\cite{moreno2023generalized} to model the response of the different adhesives under planar extension. The LHMM considers a two-way coupling between micro and macroscopic scales as illustrated in Figure \ref{fgr:modelSketch}.a. At the macroscale, the system is modeled using a continuum description of the material, whereas at the microscale, a polymeric-network model is adopted. Velocity gradients measured at the macroscale are used to define boundary conditions for microscale simulations, while the stress tensor at the macroscale is determined from the microscale simulations. As illustrated in Figure \ref{fgr:modelSketch}.a, each discrete point in the macroscale description is equipped with its own microscale simulations. In the following, we describe the discretizations used for macro and microscales. 

\subsubsection{Macroscopic model}

At the macroscales, we consider an incompressible material such that $\nabla \cdot \vb = 0$, and the momentum balance read
\begin{align}
\rho {\der \vb}/{\der t} - \nabla \cdot \bm\sigma(\vb,p) &= f,
\label{eq:momentum}
\end{align}
where $\rho$, $p$, $\vb$, and $\bm\sigma$ are the density, pressure, velocity, and stress tensor, respectively. The force term $f$ accounts for external forces acting on the system. The stress tensor $\bm\sigma$ can be expressed as $\bm\sigma = p{\bf I} + \bm\pi$, where $\bm\pi$ is the viscous stress tensor. It is customary to decompose this viscous stress into hydrodynamic and polymeric (non-hydrodynamic) contributions, such that $\bm\pi = \bm\pi^o + \bm\pi^p$. The hydrodynamic stress tensor $\bm\pi^o$ accounts for Newtonian fluid-like behavior, while the non-hydrodynamic stress tensor $\bm\pi^p$ accounts for the presence of polymer networks. Following \cite{moreno2023generalized}, we consider that the hydrodynamic contributions to the stress can be fully captured at the macroscales, whereas polymeric contributions are only accounted for at the microscales, leading to a macroscopic stress tensor of the form $\bm \sigma =  -p{\bf I} + {\bm \pi^o}({\bm {v}}, {t}) + \bar{\bm \pi}^p({\bm v'}, t')$, where the upper-bar notation indicates properties determined as an ensemble average of microscopic simulations, and the prime notation indicates properties measured at the microscales, such that
\begin{align}
    \bar{\bm \pi}^p({\bm v'}, t') = \langle{{\bm \pi'}}^{p}\rangle = {\frac{1}{\Omega'} \int_{\Omega'} {{\bm \pi'}}^{p}(r') \der \Omega'}.
    \label{eq:micromacropi}
\end{align}
The polymeric contribution is determined on-the-fly from microscale simulations of size $\Omega'$. For simplicity in the notation, in the remaining, we refer to $\bar{\bm \pi}^p({\bm v'}, t')$ as $\bar{\bm \pi}^p$. In a Lagrangian framework \citep{espanol2003smoothed}, the divergence of the total stress required for \eqref{eq:momentum} takes the form
\begin{equation}
\nabla \cdot {\bm \sigma} = \underbrace{-\nabla p + \epsilon \Big({\eta} \nabla^2 \vcon + \left({\zeta} + \frac{{\eta}}{D}\right) \nabla \nabla \cdot \vcon \Big)}_{\text{macroscopic}} + \underbrace{\nabla \cdot \bar{\bm \pi}^p}_{\text{microscopic}},
\label{eq:divstress}
\end{equation}
where $D$ is the dimension, and ${\eta}$ and ${\zeta}$ are the shear and bulk viscosities, respectively. Given \eqref{eq:divstress}, we use the smoothed dissipative particle dynamics (SDPD) method \citep{espanol2003smoothed,ellero2018everything} to discretize Eq. \eqref{eq:momentum}. SDPD is widely used in soft matter modeling \cite{Moreno2013, Nieto2022} as it provides a thermodynamically consistent method for discretizing the fluctuating Navier-Stokes equations. Moreover, as the volume ${\mathcal{V}}_I$ of the discretizing particles approaches continuum scales, the thermal fluctuations become negligible, and SDPD is equivalent to the well-known smoothed particle hydrodynamics (SPH) method \citep{vazquez2009consistent}. Hence, an adhesive is represented macroscopically as a collection of particles, each with a mass ($m_I$), volume ($\mathcal{V}_I$), position (${\bm r}_I$), velocity ($\vcon_I$), and polymeric stress tensor (${\bm \pi}_I^p$). The evolution of the particle's position is given by ${\der {\bf{r}}_I}/{\der t} = \vcon_I$, whereas from \eqref{eq:momentum} and \eqref{eq:divstress}, the particle's number density $d_i$ and the momentum equation can be expressed as 
\begin{align}
m_I\frac{\der {d}_I}{\der t} &= \sum_J \fijcon \rbijcon\cdot\vcon_{IJ}\label{eq:rhoevo}
\\
{m}_I\frac{\der \vcon_I}{\der t}  &= \sum_J  \Big[\frac{{p}_I}{d_I^2} + \frac{{p}_J}{d_J^2}\Big] \fijcon {\rbijcon} \nonumber \\ &- \sum_J \Big[\facon \vcon_{IJ} +\fbcon (\vcon_{IJ}\cdot\ebijcon)\ebijcon \Big]  \frac{\fijcon}{d_I d_J} - \sum_J \bar{\bm \pi}_{IJ}^p \fijcon {\rbijcon},
\label{eq:momsph}
\end{align}
where $\rbijcon = {\bm r}_I - {\bm r}_J$, $\vcon_{IJ} = \vcon_I - \vcon_J$, and $\ebijcon= \rbijcon/|\rbijcon|$. $\fijcon$ is a positive function defined as $\fijcon =- \nabla W(|\rbijcon|,h)/|\rbijcon|$, where $W(|\rbijcon|,h)$ is an interpolant kernel with finite support $h$ and normalized to one, as is customary in SDPD (and SPH). The term ${p}$ is the density-dependent pressure, whereas $\facon$ and $\fbcon$ are friction coefficients related to the shear ${\eta}$ and bulk ${\zeta}$ viscosities of the system through $\facon={(D+2){\eta}}/{D}-{\zeta}$ and $\fbcon = (D+2)({\zeta}+{{\eta}}/{D})$. In the last term of \ref{eq:momsph}, we have that $\bar{\bm \pi}_{IJ}^p={\bar{\bm \pi}_I^p}/{d_I^2} + {\bar{\bm \pi}_J^p}/{d_J^2}$, where the stress on particles $I$ and $J$ is computed from microscale simulations using the Irving-Kirkwood formalism\cite{Irving1950,moreno2023generalized} (see Supplementary Data S.3). 
\subsubsection*{Microscopic model}

In general, it is expected that the properties of cross-linked polymeric materials are influenced by microscopic features such as the chemical structure of the polymer chains, the cross-linking density,  and the network topology. The cross-linking density, which is the number of cross-links per unit volume, plays a crucial role in determining the material's mechanical properties. A higher cross-linking density leads to a stiffer material with higher modulus and strength, while a lower cross-linking density results in a more flexible material \cite{Rubinstein2023, abd2003effect,kumar2007allyl}. The network topology of polymer chains, defined by the arrangement and connectivity of crosslinks and entanglements, plays a critical role in determining mechanical properties, as it influences elasticity, strength, toughness, and the ability to dissipate stress under deformation \cite{Svaneborg2008}. Here, we adopt the network formation methodology introduced by Zohravi et al. \cite{zohravi2023computational} that allows us to capture the effects of cross-linking density and network topology on the material's mechanical properties. This framework accounts for the kinetics of network formation, and likewise, the macroscopic model is built into SDPD discretization. Additionally, it has been successfully used to capture rheological and structural features of polymeric systems \cite{zohravi2025mesoscale}.  

It has been demonstrated experimentally that the polymer networks synthesized in a compartmentalized polymerization process, such as emulsion polymerization, contain a high molecular weight "gel" component that is crosslinked, as well as lower molecular weight soluble chains. Additionally, the correlation length, $\xi$ of the crosslinked component, is much larger than the size of the radius of gyration $R_g$ of the non-crosslinked polymer chains \cite{Mehravar2018}. Therefore, the gel network can be represented by interconnected blobs with a characteristic size $\xi$. In this representation, the mechanical properties of the gels depend on the contributions of intra- and inter-blob interactions. Following this representation, we consider that the polymeric network systems are constituted by two types of particles, denoted gel (\ac) and sol (\sol), as illustrated in Figure ~\ref{fgr:modelSketch}.b. In this context, the sol refers to a different class of blob composed of multiple non-crosslinked (either entangled or unentangled) polymer chains

The gel volumetric fraction, $\phi_{\text{gel}}$, is defined as the ratio of the number of gel (\ac) particles to the total number of particles (gel plus sol particles). All the particles interact hydrodynamically using the standard SDPD method\cite{espanol2003smoothed}, including thermal fluctuations, characteristic at smaller scales. To model the network connectivity between \ac\ particles, we use a Morse bond potential\cite{morse1929diatomic} and define a crosslinking degree $c_d$ (maximum number of bonds per particle). Notice that these \textit{bonds} do not represent chemical bonds between atoms; instead, they indicate an effective inter-blob connectivity. This connectivity is a conceptual tool used to describe how different regions (or "blobs") of the material interact with each other. It is not a physical bond but rather a representation of the material's structural organization at a mesoscale level. We use this concept to correlate with variations in the crosslinking density of the material, which affects its mechanical and structural properties. This idea is illustrated in Figure ~\ref{fgr:modelSketch}.c for two different values of $c_d$, where $c_d$ represents the crosslinking density. By varying $c_d$, we can observe how changes in the effective connectivity influence the material's behavior. In our modelling strategy, we do not explicitly resolve intra-blob interactions but consider each \ac\ particle as a coarse-grained polymer-blob forming the network, and the particles of type \sol\ represent lower molecular weight polymer chains that are not crosslinked into the gel structure. Thus, the strength of the bonding potential and the maximum number of bonds per particle are adjusted to reproduce the effective mechanical properties of the material\cite{zohravi2025mesoscale}. Note that this framework does not consider any spatial heterogeneity of the gel and soluble fractions that make occur in the polymer film as a result of the drying process from the initial polymer dispersion.

Similar to macroscales, when using SDPD, we consider that the microscopic polymeric systems are discretized as a set of interacting particles with mass $m_i'$, volume $\mathcal{V}_i'$, position ${\bf r}_i'$, and velocity $\vmes_i$. For clarity, we use lower case indices $i$ and $j$ to refer to particles at the microscale, in addition to the prime notation indicating microscale variables (e.g, $\vmes_i$). The position's evolution of particles at the microscale is given by ${\der {\bf{r'}}_i}/{\der t} = \vmes_i$, whereas the momentum equation can be expressed as 
\begin{align}
m_i'\frac{\der \vmes_i}{\der t}  = &\sum_j \left[ \frac{p'_i}{d_i^2} + \frac{p'_j}{d_j^2}\right] \fij \rbij \nonumber \\ &-\sum_j \left[\fa \vmes_{ij} +\fb (\vmes_{ij}\cdot\eij)\eij \right] \frac{\fij}{d_id_j} \nonumber \\
&+\sum_j \left(\Aij \der \wbbarij + \Bij \frac{1}{D}\text{tr}[ \der \wbij] \right) \cdot \frac{\eij}{\der t} + \sum_j \kappa_{ij}F_{ij}^{\text{bond}},
\label{eq:deterministic}
\end{align}
where $\vmes_i$ and $p'_i$ are velocity and pressure of the $i$-th particle. $\vmes_{ij} = \vmes_i - \vmes_j$, $\eij = \rbij/|\rbij|$, $\fa$ and $\fb$ are friction coefficients related to the shear $\eta'$ and bulk $\zeta'$ viscosities of the fluid through $\fa={(D+2)\eta'}/{D}-\zeta'$ and $\fb = (D+2)(\zeta'+{\eta'}/{D})$. Equivalent to \eqref{eq:momsph}, in \eqref{eq:deterministic} we have that $F'_{i j} =- \nabla W(\rbij,h')/|\rbij|$. The third term in equation \eqref{eq:deterministic}, is thermodynamically consistent and accounts for thermal fluctuations at microscales as described by Espa\~nol and Revenga\cite{espanol2003smoothed}. Where $\wbbarij$ is a matrix of independent increments of a Wiener process for each pair $i,j$ of particles, and $\wbbarij$ is its traceless symmetric part, given by $\der \wbbarij = {1}/{2}\left[d\wbij+\der \wbij^T\right] - {\delta^{\alpha \beta}}/{D}\text{tr} [ \der \wbij].$ The thermal noises $\Aij$ and $\Bij$ are related to the friction coefficients $\fa$ and $\fb$ through $A_{ij}= \left[4\kbt \fa {\fij}/{\rhoi \rhoj} \right]^{1/2}$ and $B_{ij}= \left[4\kbt\left(\fb -(\fa{D-2})/{D}\right){\fij}/{\rhoi \rhoj} \right]^{1/2}$
where $\kbt$ is the thermal energy, and $\rhoi$ and $\rhoj$ are the densities of the particles $i$ and $j$. The pressure $p'_i$ of the $i$-th particle is given by the equation of state of the form  $p'_i = {c^2\rho_0}/{7}\left[ ({\rho_i}/{\rho_0})^{7}-1\right]$. Where $c$ is the artificial speed of sound on the fluid, and $\rho_0$ is the reference density. In \ref{eq:deterministic}, the term $\kappa_{ij}=1$  if the pair $i$-$j$ is connected or zero otherwise. Following the methodology of Zohravi et al\cite{zohravi2025mesoscale}, $F_{ij}^{\text{bond}}$ accounts for the force between connected particles and is given by \cite{morse1929diatomic}
\begin{equation}
F_{ij}^{\text{bond}}= 2K_s[e^{-2(r_{ij}-r_0)}-e^{-(r_{ij}-r_0)}]
\label{eq:bond}
\end{equation}
where $r_0$ and $K_s$ are the equilibrium bond distance and stiffness constant, respectively. Although the polymer network connectivity is described by a Morse-like bond potential \eqref{eq:bond}, the overall system exhibits viscoelastic behavior due to the embedding of the bonded network within the viscous medium generated by the SDPD method. Elasticity arises from bond stretching and compression, while viscosity originates from SDPD interactions both between solvent and bonded particles and among bonded particles themselves (Figure~\ref{fgr:modelSketch}.b). The viscous forces in SDPD depend on relative particle velocities, and their coupling with the elastic bond network yields an effective viscoelastic response of the system.

In \eqref{eq:bond}, we fixed $r_0$, and conducted a parametric study for different polymeric networks varying the bond stiffness $K_s$ and the maximum number of bonds per particle -- or crosslinking degree -- $c_d = 2-4$  to reproduce the experimental variations in $G'$ and $G''$ for the four adhesive samples. We performed SAOS \cite{zohravi2025mesoscale} characterization to determine the viscous and loss modulus of the simulated gel. For each sample, the volume fraction $\phi_{\text{gel}}$ of gel particles is prescribed from the experimentally measured gel content. In Supplementary Data S2, we present a detailed description of the stages adopted for gel formation and the methodology used to calibrate the microscale model. The parameters in our microscale simulations are referred to the characteristic length ($L_{\text{sdpd}}$), mass ($m_{\text{sdpd}}$), and time ($\tau_{\text{sdpd}}$) units. 

\subsubsection*{Coupling between scales and mapping}

Given the macroscopic \eqref{eq:momsph} and microscopic \eqref{eq:deterministic}  evolution equations, the information exchange occurs at every time $t_{\text{coupling}}$. In LHMM, the macro-to-micro coupling is achieved by imposing the velocity gradient $\nabla \vcon_I$, measured on the $I$th-particle, as the boundary condition for its microscale simulation\cite{Moreno2021} (see Figure \ref{fgr:modelSketch}.a). Microscale simulations evolve according to \eqref{eq:deterministic} for a number of timesteps $n\Delta t'=t_{\text{coupling}}$, during which the mean stress tensor $\bar{\bm \pi}_I^p$ is determined using the Irving-Kirkwood formalism\cite{Irving1950, Thompson2009}. The micro-to-macro coupling is done using $\bar{\bm \pi}_I^p$ to evolve the macroscale simulations for $N\Delta t = t_{\text{coupling}}$ according to \eqref{eq:momsph}. Notice that despite both scales can have different time step sizes (i.e., $\Delta t$ and $\Delta t'$), the coupling time is always the same, such that the scales are executed synchronously.  See Supplementary Data S3 for a detailed description of the methodology to compute the velocity gradient and viscous stress tensor.  

In general, the numerical parameters (e.g., viscosity, density, etc) used for both scales in \eqref{eq:momsph} and \eqref{eq:deterministic} can be made non-dimensional by conveniently defining characteristic scales (e.g., time $[\tau_{\text{sdpd}}]$, length $[L_{\text{sdpd}}]$, mass $[m_{\text{sdpd}}]$). This approach is customary for numerical purposes to avoid too large (or small) numbers in the simulations. If we consider the zero shear polymeric viscosity $\eta_p$ (and $\eta_p'$) and the maximum shear rate $\gamma_{\text{max}}$ (and $\gamma_{\text{max}}'$) at each scale, we can express a set of dimensionless stress tensors of the form $\bar{\bm \pi}^p/{\eta_p \gamma_{\text{max}}}$ and ${\bm \pi'}^p/{\eta_p' \gamma_{\text{max}}'}$. This effectively means that both scales can be mapped into physical systems using different scaling factors. However, to ensure the correct coupling between scales, the dimensionless stress tensors must be equivalent in both scales ($\bar{\bm \pi}^p/{\eta_p \gamma_{\text{max}}}={\bm \pi'}^p/{\eta_p' \gamma_{\text{max}}'}$). Therefore, \eqref{eq:micromacropi} can be rewritten as
\begin{align}
    \bar{\bm \pi}^p({\bm v'}, t') = \alpha \tau_{m} \langle{{\bm \pi'}}^{p}\rangle
    \label{eq:micromacropi2}
\end{align}
where the parameter $\alpha = {\eta_p}/{\eta_p'}$ indicates the ratio between the polymeric viscosities simulated at each scale, whereas $\tau_m = {\gamma_{\text{max}}}/{\gamma_{\text{max}}'}$ indicates a ratio between characteristic times at each scale. Notice that for macroscales, $\eta_{\text{total}}= \eta + \eta_p$, where $\eta$ is the  fluid viscosity in \eqref{eq:momsph}, and $\eta_p$ corresponds to the polymeric contribution to the total viscosity. The latter defines a viscosity ratio $\beta$ given by
\begin{equation}
    \beta = \frac{\eta_p}{\eta_{\text{total}}} = \frac{\alpha\eta_p'}{\eta + \alpha\eta_p'},
    \label{eq:beta}
\end{equation}
where the parameter $\beta$ can be adjusted for macroscopic simulations to reproduce the experimental measurements, by varying the fluid viscosity $\eta$ and $\alpha$, whereas $\eta_p'$ is prescribed by the conditions of the microscale simulations. For consistency, the velocity gradient $\nabla \vcon_I$ (measured at macroscales) defining the boundary conditions for the microscale simulations needs to be rescaled by $\tau_{m}$ to ensure both scales remain synchronized, such that $\nabla \vcon_I' = \nabla \vcon_I/\tau_{m}$, where $\nabla \vcon_I'$ is the actual velocity gradient imposed at microscales. 

\subsubsection*{Tensile test simulation setup}

We conducted a series of 2D simulations to investigate the mechanical response of the different formulations under uniaxial tensile loading, as schematically depicted in Figure \ref{fgr:modelSketch}.a. In a experimental tensile test, a freestanding or clamped polymeric film is subjected to a controlled elongation, and the resulting tensile force is recorded as a function of the applied strain. In the present simulations, the polymer film is confined between two parallel solid walls, with the upper wall displaced at a constant velocity to impose a uniform extensional deformation. The tensile force acting on the moving wall is measured and used to construct stress–strain curves for each formulation. We ensure non-slip boundary condition at the interface between the adhesive material and the walls following Morris et al\cite{Morris1997, Bian2012}, to warrant that the  sample is always attached to the wall. 

Experimentally, the tensile stress–strain response of polymeric films typically comprises three distinct regimes: \textit{i}) an initial linear elastic region governed by entropic chain elasticity, \textit{ii}) a nonlinear or yield regime associated with the onset of plastic deformation, chain alignment, and entanglement rearrangement, and \textit{iii}) a failure regime characterized by strain localization, necking, or rupture. It should be noted that the present model provides a purely two-dimensional representation of the polymeric film, whereas in real tensile experiments the film thickness evolves during deformation. Consequently, the simulations constitute an approximation of the experimental system, and quantitative differences may arise, particularly in the maximum strain attained prior to failure. Despite these limitations, the overall tensile response remains a stringent benchmark for assessing the capability of the multiscale modeling framework to capture the mechanical behavior of polymeric adhesives across length scales. To enable direct comparison with experimental tensile data, the strain is normalized by the maximum strain, $\epsilon^*$, measured for the reference (AD1) sample. This approach facilitates systematic comparison among formulations and provides insight into the molecular mechanisms underlying their distinct mechanical responses.

Given that the microscale response of the polymeric film is represented by a mesoscale network of SDPD particles interacting via \eqref{eq:bond}, with a characteristic equilibrium length, in LHMM, we introduce the bond-breaking distance $r_c$ as a numerical and physical regularization parameter that defines the maximum admissible bond extension (or equivalently, the stress threshold) beyond which a bond is irreversibly removed. Allowing bonds to break is essential to prevent numerical instabilities under large deformations in multiscale simulations and to enable the microscale model to accommodate irreversible structural rearrangements induced by macroscopic loading.

From a modeling standpoint, the $r_c$ acts as an effective material parameter that controls the onset of network degradation and strongly influences the overall mechanical response, in conjunction with macroscopic viscosity $\eta$ and the polymer viscosity ratio $\beta$ (see \eqref{eq:beta}). Physically, this parameter can be interpreted as a coarse-grained proxy for the combined effects of crosslink density and polymer entanglement strength, both of which govern the extensibility and load-bearing capacity of the network. While alternative formulations could explicitly distinguish between permanent crosslinks and transient entanglements through multiple bond types or double-network-like gel models \cite{Walker2025, Mugnai2025}, the present model employs a single bond population to capture their combined mechanical contribution. As a result, $r_c$ requires material-specific calibration, but provides a straightforward and effective means of capturing the mesoscale response to reproduce experimentally observed tensile behavior within the multiscale framework.

In Supplementary Table S.3, we summarize the values used for the macro and microscale simulations. The parameters are chosen to ensure that the simulations are consistent with the experimental conditions. Macroscopic fluid viscosity $\eta$, the viscosity ratio $\beta = \eta_p/\eta_{\text{total}}$, and the bond-breaking distance of the microscopic simulations are varied to reproduce the tensile experiments. 

\section*{Results}

\subsection{Experimental Characterization of Adhesive Formulations} 
\begin{figure*}[!t]
    \centering
    \includegraphics[width=1\linewidth]{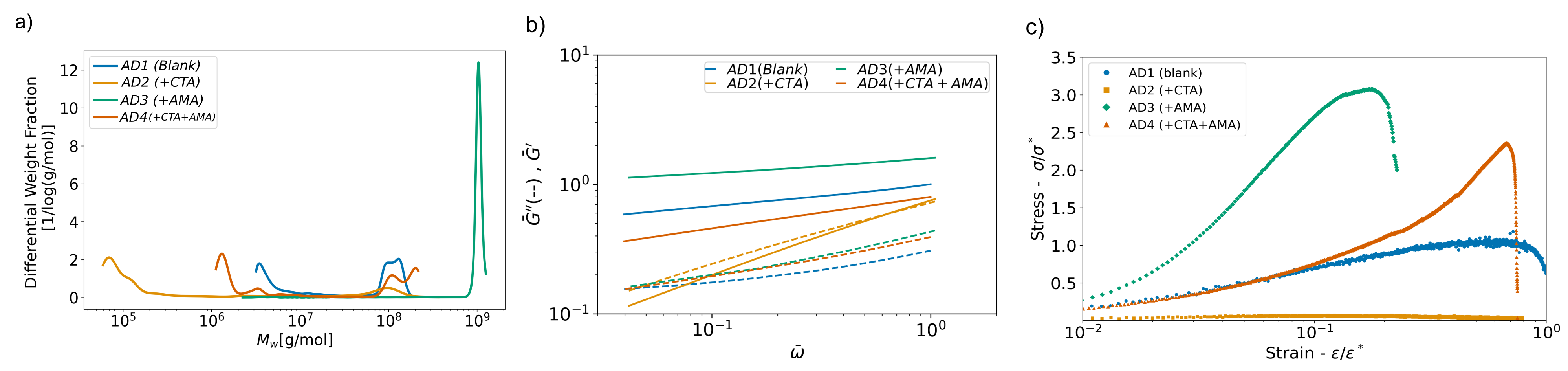}
    \caption{Experimental characterization of the different samples. a) Differential molecular weight 
    fraction, b) SAOS rheological characterization. Continuous and dashed lines corresponds to experimental measurements of $G^\prime$ and $G^{\prime\prime}$,  c) Normalized stress vs strain  tensile experiments}
    \label{fgr:expSumm}
\end{figure*}

In Figure \ref{fgr:expSumm}, we summarize the experimental characterization of the different adhesive formulations. Figure \ref{fgr:expSumm}.a, presents the complete molar mass distribution for different samples measured by AF4.  It can be seen that the sample AD1, which is the pure butyl acrylate sample, exhibits two distinct peaks in the molar mass distribution. The peak at lower molar mass corresponds to the soluble polymer fraction, while the peak at higher molar mass corresponds to the macroscopic gel fraction, measured to be 69\% for this sample (Table \ref{tab:ad_formulation}). It is worth mentioning that during the polymerization of acrylates, even in the absence of a crosslinker, a polymer network can form via an intermolecular-chain transfer to polymer mechanism, followed by termination by combination \cite{Ballard2018b, Plessis2000}. In this process, the growing radical abstracts a labile hydrogen atom from another polymer chain, leading to branching and network formation. In sample AD2, the addition of a chain transfer agent (CTA) to the formulation reduces the kinetic-chain length of the polymer coils, leading to a shift of the sol polymer peak to lower molar masses, along with a significant reduction in the high-molar-mass peak and a corresponding decrease in the gel fraction to 15\%. In sample AD3, the crosslinker (AMA) added to the formulation led to further linking of the polymer chains. As a result, no peak at low molar mass was observed in the AF4 analysis. A unimodal molar mass distribution, corresponding to the gel polymer, was observed, and this distribution was shifted toward higher molar masses. The macroscopic gel content was measured to be 94\%. In contrast, sample AD4, which contains both a CTA and AMA, exhibits two peaks. The final gel content is similar to that of sample AD1 (see Table \ref{tab:ad_formulation}). However, compared to AD1, the peak corresponding to the sol polymer is shifted to lower molar masses due to the CTA’s effect in reducing the kinetic chain length \cite{Chauvet2005}. Additionally, the gel polymer peak is shifted to higher molar masses, indicating enhanced chain crosslinking in the presence of the crosslinker. 
\\\\
Figure \ref{fgr:expSumm}.b, shows the experimental storage and loss modulus of the different samples. Consistent, with their molar mass distribution, we observe a characteristic behavior of higher storage modulus for samples with higher gel content, such that $G_{\text{AD3}}'>G_{\text{AD1}}'>G_{\text{AD4}}'>G_{\text{AD2}}'$. Moreover, as crosslink density increases, the decrease in storage modulus at low frequencies becomes less pronounced, reflecting the enhanced elastic behavior of the network. 

Figure \ref{fgr:expSumm}.c illustrates the tensile stress response of the different adhesive formulations, highlighting clear differences in their deformation and failure behavior. The blank sample AD1 exhibits a gradual increase in stress at the onset of deformation, corresponding to its elastic response, followed by a relatively sustained stress level prior to failure. This plateau-like behavior indicates moderate resistance to deformation before rupture.

In contrast, adhesive AD2 shows minimal stress buildup during elongation, reflecting its tendency to deform with little mechanical resistance. This response is consistent with its lower molecular weight and its liquid-like rheological behavior, as indicated by the dominance of the loss modulus over the storage modulus. At the opposite end of the spectrum, the densely crosslinked adhesive AD3 displays a high onset stress that increases rapidly with deformation, leading to abrupt fracture. This behavior is characteristic of a highly elastic network in which reduced interchain mobility results in stress localization and sudden failure once the intermolecular crosslink strength is exceeded.

The response of sample AD4, which contains both AMA and CTA, lies between these limiting cases and reflects a balance between sol and gel fractions. At small deformations, AD4 resembles AD1, exhibiting a gradual stress increase. However, at larger strains, the stress rises sharply, similar to AD3, culminating in fracture. Notably, this failure occurs at significantly larger deformations than in the AMA-only formulation (AD3), indicating an intermediate mechanical response governed by its mixed network structure.

\subsection{Computational Rheological Characterization of Microscales}

\begin{figure}[!thbp]
\centering
\includegraphics[width=1\linewidth]{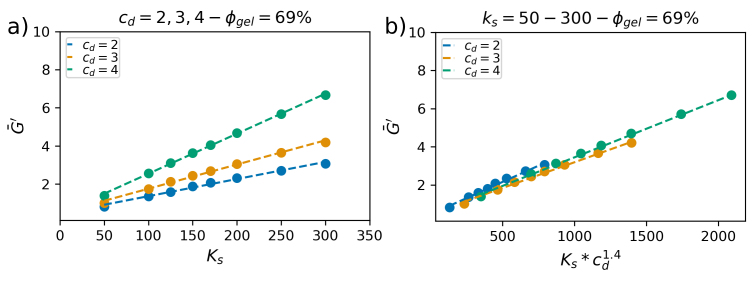}
\caption{a) $G^{\prime}$ linear correlation with $K_s$ and b) exponential correlation with $c_d$, $G^{\prime} \sim K_s{c_d}^{1.4}$. $G^{\prime}$ measured at maximum frequency ($\bar{\omega}=1$) at SAOS test for $\phi_{gel}=69\%$ and normalized with the corresponding value at $ K_s=50$ and $c_d=2$.}
\label{fgr:Gprimks}
\end{figure}

Based on the experimental characterization results, we constructed computational microscale models for the different samples, using their corresponding gel content (see Table \ref{tab:ad_formulation}) to define $\phi_{\text{gel}}$ in the simulations. Then, we systematically tuned the cross-linking degree $c_d$ and bond stiffness $K_s$ to reproduce the experimental slope of $\bar{G}^{\prime}$ as a function of $\bar{\omega}$, and the $G^{\prime\prime}/G^{\prime}$ ratio (see Supplementary Data S2.3 for a detailed description on the mapping between numerical to experimental parameters). Given that the increase in the cross-linking density should lead to higher storage moduli (corresponding to a more accentuated elastic behavior), we aimed to identify a numerical correlation that allow us to map experimental rheological measurements to cross-linking densities in our computational model\cite{abd2003effect,kumar2007allyl}.  In general, for the different cross-link degree $c_d$ simulated, we identified a linear relationship of $G^{\prime}$ with the bond stiffness $K_s$, as illustrated in Fig.~\ref{fgr:Gprimks}.a for the gels with $\phi_{\text{gel}}=69\%$, varying $c_d  = 2-4$, at $\bar{\omega}=1$. In Fig.~\ref{fgr:Gprimks}.a, we also observed that a larger cross-link degree effectively led to an increase in $G^{\prime}$, due to the higher interconnected network. This dependence of $G^{\prime}$ on $c_d$ seems to follow a power-law behavior. From our simulation results, we identified that the elastic modulus, up to a good approximation, correlated with $G^{\prime} \sim K_s c_d^{1.4}$, as illustrated in Fig.~\ref{fgr:Gprimks}.b. This correlation served as a proxy (or mapping) to determine the model parameters $K_s$ and $c_d$ that best reproduce the $G^{\prime}$ dependence for experimental samples.  

\begin{figure}[!thbp]
\centering
\includegraphics[width=1\linewidth]{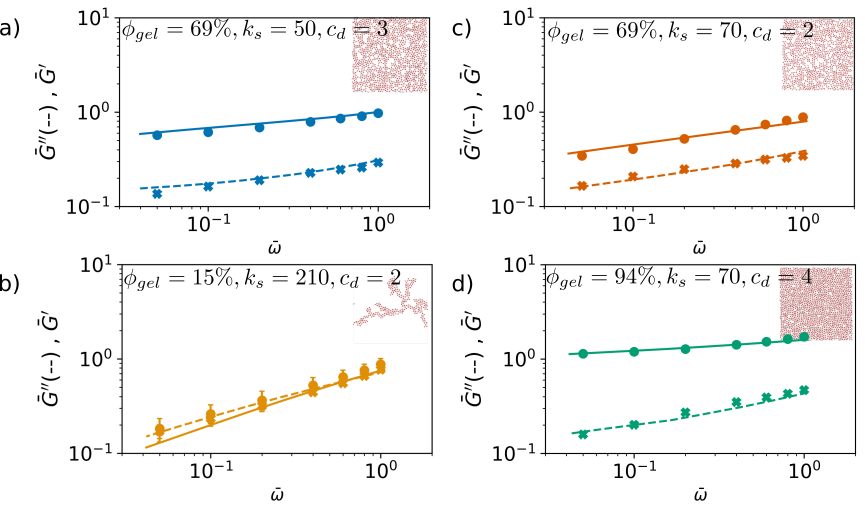}
\caption{$G^{\prime}$ and $G^{\prime\prime}$ as a function of frequency for experimental (lines) and numerical ($\bullet,\times$) results for the different adhesive formulations. The values of $\phi_{\text{gel}}$, $K_s$, and $c_d$ used to simulate the gels are indicated. a) AD1, b)AD2, c)AD4, d)AD3. A visualization of the \ac\ particles forming the network is shown for each sample. Particles of type \sol\ are omitted for clarity. }
\label{fgr:expnum}
\end{figure}

Following the identified mapping of the numerical parameters ($K_s$ and $c_d$), we constructed  four different gels that reproduce the response identified for AD1 to AD4.  In Figure \ref{fgr:expnum}, we present the comparison of the simulated and experimental modulus. Overall, we found a satisfactory agreement in dependence of $G^{\prime}$ and $G^{\prime\prime}$ for the different samples. In Figure \ref{fgr:expnum}, we indicate the values of $\phi_{\text{gel}}$, $K_s$, and $c_d$ used to model the gels, along with an example of the structure of the network. Consistent with the experimental results, the cross-linking degree $c_d=4$ for sample AD3 correlated with the largest cross-linking degree expected for formulations using AMA. Similarly, the lowest value $c_d=2$ for sample AD2 evidenced the reduction in the crosslinking induced by the use of CTA. Remarkably, the polymeric network formation methodology used allows us to capture the proper rheological response of the different samples while providing insights into the network topology and cross-linking density. It may be noted that other adhesive formulations with alternative compositions can be easily mapped into this numerical model to investigate their multiscale response.

\subsection{Tensile Experiments}

Once the rheological response of the microscales was properly calibrated, we proceeded to simulate the tensile experiments for the different samples. As described in the methodology, each macroscopic particle discretizing the material is equipped with a microscale simulation containing the polymeric network. In principle, to properly represent the microstructural variety at the different locations of the sample, each of the microscale simulations should correspond to a statistically independent realization of the network. However, for the sake of computational efficiency, we consider five independent realizations of the networks (with constant $K_s$ and $c_d$) that are randomly distributed in the whole macroscopic domain. Thus, each of the macroscopic particles is equipped with one out of five gel structures. 

\begin{figure*}[!t]
\centering
\includegraphics[width=.9\linewidth]{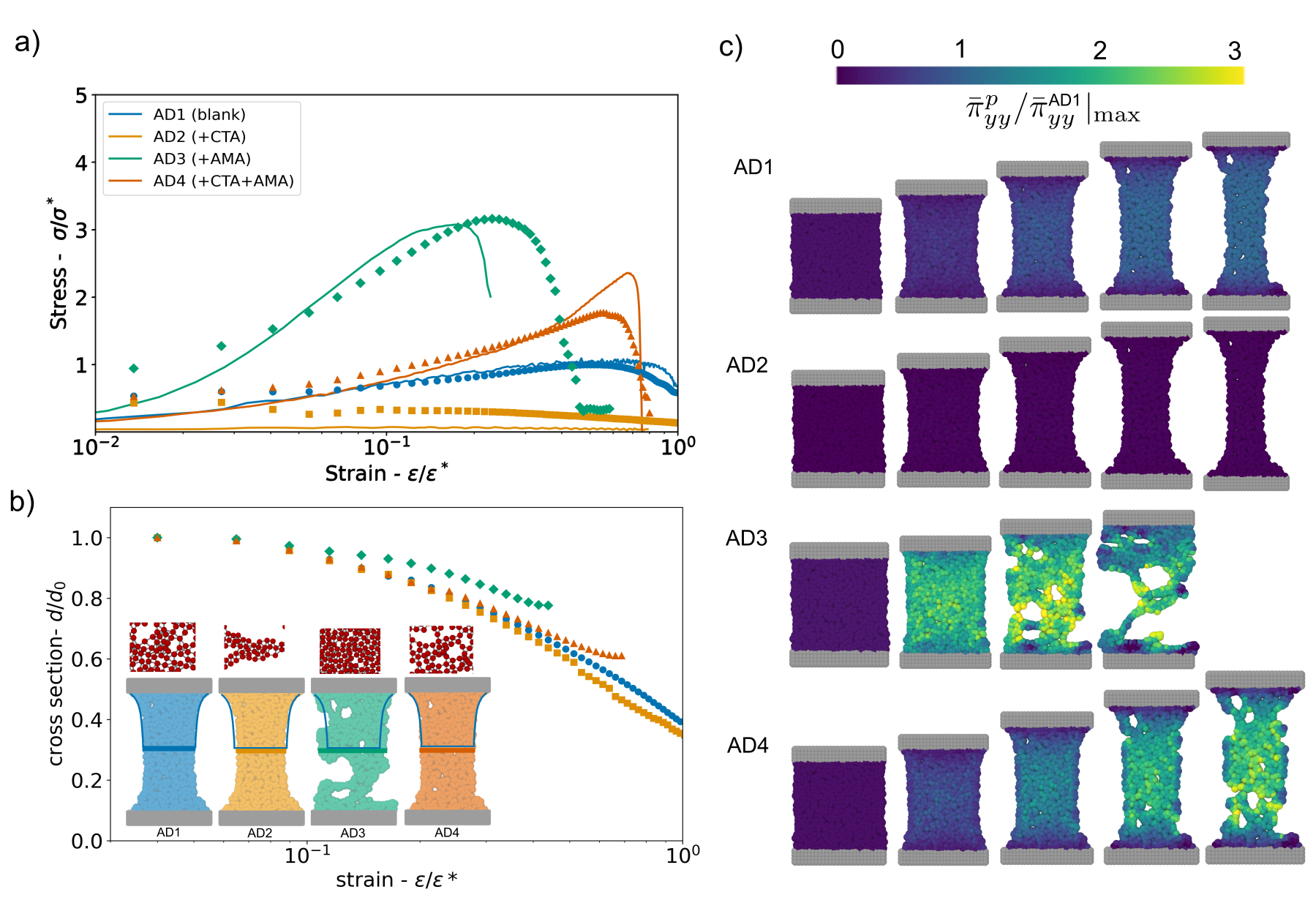}
\caption{a) Comparison of experimental (solid line) and numerical (solid markers) results for tensile experiments for the four different adhesive formulations. Corresponding to macroscopic conditions for AD1 to AD4 with $\eta=100$, $\beta = \{0.68, 0.22, 0.9, 0.6\}$, and $r_c = \{1.8 ,1.3, 1.4, 2.4\}r_0$, where $r_0$ is the equilibrium bond distance at the microscales.  b) Variation in the cross section width $d/d_0$, as a function of strain. Inside snapshots illustrate the microstructure of each adhesive and their corresponding macroscopic morphology in the tensile test at $\epsilon/\epsilon^*=0.4$ strain. The width at the center of the samples is highlighted with bold lines. The profile of the stretched AD1 sample is added as an overlay over all the adhesives for comparison. c) Microscopic stress $\bar{\pi}^p_{yy}$ evolution for the different samples. The colorbar corresponds to $\bar{\pi}^p_{yy}/\bar{\pi}^{\text{AD1}}_{yy}|_{\text{max}}$, where $\bar{\pi}^{\text{AD1}}_{yy}|_{\text{max}}$ is the maximum stress measured for AD1 sample during the tensile test.}
\label{fgr:numTack}
\end{figure*}

Figure~\ref{fgr:numTack} summarizes the numerical results of the tensile simulations. Figure~\ref{fgr:numTack}.a presents a direct comparison between the simulated stress–strain curves and the corresponding experimental data for all formulations. Overall, the simulations reproduced the main qualitative features of the experimental response, including the initial low-strain behavior, the approximate location of the peak stress, and the subsequent stress relaxation or fracture.

Minor discrepancies are observed in the predicted onset stress, in the position of the maximum stress for AD3, and in the peak stress magnitude for AD4. Nevertheless, the model successfully captures the distinct mechanical signatures of each formulation, which represents a significant outcome of the multiscale framework. While further refinement and calibration of model parameters could improve quantitative agreement, the ability to differentiate the tensile responses of the various samples already demonstrates the predictive potential of the approach. In particular, this capability opens the possibility of using the model as a tool to screen and guide adhesive formulation design by linking macroscopic performance to underlying network characteristics.

Figure~\ref{fgr:numTack}.b presents the evolution of the normalized cross-section width, $d/d_0$, where $d$ and $d_0$ denote the instantaneous and initial cross-section width, respectively. To facilitate comparison, the macroscopic morphologies of all samples are shown at a representative strain of $\epsilon/\epsilon^* \simeq 0.4$, where $\epsilon^*$ is the maximum strain reached by the reference sample AD1. The deformed shape of AD1 is superposed onto the other samples to emphasize differences in lateral contraction and structural integrity.

Clear trends emerge as a function of network architecture. The highly crosslinked formulation AD3 maintains a comparatively rigid structure, exhibiting limited width reduction prior to abrupt, catastrophic failure. In contrast, AD2—characterized by a low crosslinking density and more fluid-like rheological behavior—undergoes pronounced and nearly monotonic lateral contraction, resulting in a significantly narrower cross-section. AD4 displays an intermediate response: at small strains, its width contraction resembles that of AD1 and AD2, indicating appreciable chain mobility; however, at larger deformations, the reduction in width becomes more constrained, similar to AD3, preceding fracture. This progression reflects the combined influence of sol and gel fractions in its network structure.

To further elucidate the link between microstructure and macroscopic mechanical response, Figure~\ref{fgr:numTack}.c shows the spatial distribution of the tensile stress component, $\bar{\tau}_{yy}^p$ at five representative strain levels. As expected, AD3 develops higher local stress concentrations throughout the sample, consistent with its dense crosslinked network and limited stress relaxation capacity. By contrast, AD1, AD2, and AD4 exhibit more homogeneous stress distributions at comparable strains. Notably, in AD4, the later stages of deformation reveal the onset of localized stress amplification and incipient crack formation, accompanied by a marked increase in stress heterogeneity. These observations highlight how differences in crosslink density and network connectivity govern both macroscopic deformation patterns and the spatial organization of internal stresses prior to failure.

\section*{Conclusions} 

In this work, we developed and validated a multiscale computational framework for predicting the rheological and mechanical behavior of pressure-sensitive adhesives (PSAs). The approach is based on the Lagrangian Heterogeneous Multiscale Method (LHMM), which enables a consistent coupling between macroscopic continuum deformation and a mesoscale network representation of the polymer microstructure. Model parameters were calibrated against experimental rheology and tensile data for four adhesive formulations with different crosslinking densities and gel fractions, allowing for a systematic assessment of structure–property relationships across scales.

The simulations reproduce the main experimental trends in storage modulus ($G'$), loss modulus ($G^{\prime\prime}$), and tensile stress–strain response. In particular, the framework captures how variations in crosslink density and effective bond strength govern the balance between elasticity and viscous dissipation. Formulations with higher crosslink density exhibit increased stiffness, elevated stress concentrations, and limited capacity for stress redistribution, leading to earlier catastrophic failure under tension. In contrast, weakly crosslinked systems display enhanced chain mobility, more homogeneous stress fields, and greater deformability prior to fracture. These results demonstrate that the proposed framework can resolve the interplay between network topology, entanglement constraints, and macroscopic mechanical performance, thereby providing a mechanistic interpretation of experimentally observed trends.

Although the present model successfully describes bulk rheological and tensile behavior, several aspects of PSA performance remain to be incorporated. In particular, adhesive–substrate interactions, cavitation, fibrillation, and interfacial debonding, phenomena central to classical probe-tack experiments, are not explicitly resolved in the current formulation. Furthermore, the two-dimensional representation of the film and the effective treatment of bond rupture introduce simplifying assumptions that may influence quantitative predictions at large strains. Future developments will therefore focus on extending the macroscopic description to include interfacial mechanics, damage evolution, and three-dimensional effects, as well as refining the multiscale coupling strategy.

Despite these limitations, the proposed LHMM-based framework provides a robust and computationally tractable platform for linking molecular-scale design parameters to macroscopic adhesive performance. Its ability to differentiate formulations and reproduce experimental trends highlights its potential as a predictive tool for formulation screening and rational design of next-generation PSAs.

\section*{CRediT authorship contribution statement}

\textbf{Nicolas Moreno}: Writing – review \& editing, Writing – original
draft, Visualization, Validation, Investigation, Formal analysis, Conceptualization, Resources. \textbf{Elnaz Zohravi}: Writing – review \& editing,  Writing – original draft, Investigation, Formal analysis. \textbf{Shaghayegh Hamzehlou}: Investigation, Writing – review \& editing \textbf{Edgar Pati\~no-Nari\~no}: Validation, Investigation. \textbf{Malavika Raj}: Validation, Investigation. \textbf{Mercedes Fernandez}: Validation, Investigation. 
\textbf{Nicholas Ballard}: Writing – review \& editing, Formal analysis,
Conceptualization. \textbf{Jose M. Asua}: Conceptualization, Supervision, Resources. \textbf{Marco Ellero}: Conceptualization, Supervision, Resources.

\section*{Declaration of competing interest}
The authors declare that they have no known competing financial
interests or personal relationships that could have appeared to influence
the work reported in this paper.

\section*{Acknowledgments}

The authors acknowledge the funding provided by IKUR–HPC\&AI – (HPCAI10: MOLD-POLY) and the
IKUR Strategy funded by Basque Government and the European Union NextGenerationEU/PRTR. The
research is also partially funded by the Spanish State Research Agency through BCAM Severo Ochoa
excellence accreditation CEX2021-0011 42-S/MICIN/AEI/10.13039/501100011033, and through the project PID2024-158994OB-C42 ('Multiscale Modeling of Friction, Lubrication, and Viscoelasticity in Particle Suspensions' and acronym 'MMFLVPS') funded by MICIU/AEI/10.13039/501100011033 and cofunded by the European Union. The authors thankfully acknowledges the computer resources at MareNostrum and the technical support provided by Barcelona Supercomputing Center (IM-2024-3-0013 and IM-2025-2-0042).

\bibliographystyle{elsarticle-num-names} 

\appendix

\section{Supplementary Data}

Supplementary data to this article can be found online at

\end{document}